# Cluster algorithms for anisotropic quantum spin models

Naoki Kawashima*

*Center for Nonlinear Studies and Theoretical Division*
*Los Alamos National Laboratory, Los Alamos, NM 87545*

(March 16, 1995)

## Abstract

We present cluster Monte Carlo algorithms for the $XYZ$ quantum spin models. In the special case of $S = 1/2$, the new algorithm can be viewed as a cluster algorithm for the 8-vertex model. As an example, we study the $S = 1/2$ $XY$ model in two dimensions with a representation in which the quantization axis lies in the easy plane. We find that the numerical autocorrelation time for the cluster algorithm remains of the order of unity and does not show any significant dependence on the temperature, the system size, or the Trotter number. On the other hand, the autocorrelation time for the conventional algorithm strongly depends on these parameters and can be very large. The use of improved estimators for thermodynamic averages further enhances the efficiency of the new algorithms.

Typeset using REVTEX

---

*The present address: Department of Physics, University of Tokyo, Hongo 7-3-1, Bunkyo, Tokyo 113, Japan

# I. INTRODUCTION

Recently, researchers have emphasized the importance of global updating in Monte Carlo methods. Especially for the critical phenomena and low-temperature behavior of models for condensed matter, the difficulty due to diverging numerical auto-correlation times is universal. A cluster algorithm based on the Fortuin-Kasteleyn (FK) [1] percolation representation was proposed by Swendsen and Wang [2], and was proven to be very powerful in reducing the autocorrelation time. This cluster algorithm is a successful example of a global updating Monte Carlo method. However, the Swendsen-Wang (SW) algorithm can apply only to the Ising model and its generalization is not straight-forward.

In our previous paper [3], which we will refer to as I, we generalized the FK representation to the $XXZ$ quantum spin systems. We also demonstrated [4] that the new cluster algorithm can be faster than the conventional algorithm by several orders of magnitude. In the special case of spin models with $S = 1/2$, the new algorithms are equivalent to the cluster algorithms for the 6-vertex model proposed in [5] because the $S = 1/2$ quantum $XXZ$ model can be mapped to a 6-vertex model. A similar cluster algorithm was developed also for the Hubbard model [6]. It is shown in I that the new cluster representation analogous to the FK representation is available to any model described by the $XXZ$ Hamiltonian regardless of the magnitude of the spins.

In this paper, we discuss the details of the FK-type representation and cluster algorithms for the quantum spin models which were omitted in I, namely, the model without the rotational symmetry with respect to the quantization axis. Therefore, this paper, together with I, completes the generalization of the SW-type cluster algorithm to the most general quantum spin Hamiltonian, i.e., the $XYZ$ model Hamiltonian.

In order to show the potential efficiency of the new algorithm, we performed simulations of the $XY$ model taking the quantization axis in the easy-plane.

# II. THE OUTLINE OF THE SIMULATION

The $XYZ$ Hamiltonian we consider is

$$\mathcal{H} = -\sum_{(i,j)} (J_0 + J_x S_i^x S_j^x + J_y S_i^y S_j^y + J_z S_i^z S_j^z) \quad (\boldsymbol{S}_i^2 = S(S+1)). \tag{2.1}$$

We do not assume here any special geometrical feature for the underlying lattice. The discussion given below holds for any lattice. Although the constant $J_0$ is physically irrelevant, we include it to make the subsequent discussion look simpler. Here, we assumed that the coupling constants do not depend on the sites $i$ or $j$. The generalization to inhomogeneous cases, however, is straight-forward. If $J_x = J_y$, we call the model a $XXZ$ model. The algorithms for the $XYZ$ models described in this paper is essentially the same as the ones for $XXZ$ model in I, except that we use different ways (i.e., local graphs) for breaking-up plaquettes and different probabilities for the graph assignment (i.e., labeling probabilities). For the completeness, we will briefly repeat the mathematical background of the simulation.

First, we "divide" each spin whose magnitude is $S$ into $2S$ Pauli matrices:

$$S_i^\alpha = \frac{1}{2} \sum_{\mu=1}^{2S} \sigma_{i,\mu}^\alpha \quad (\alpha = x, y, z). \tag{2.2}$$

Note that we are expanding the Hilbert space by replacing the original $2S + 1$ dimensional spin space for each spin by the $2^{2S}$ dimensional space for $2S$ Pauli spins each of which corresponds to a spin of the magnitude $1/2$. In particular, for most states in the new Hilbert space, $\boldsymbol{S}_i^2 \neq S(S+1)$ at some lattice point $i$. Therefore, such states should be excluded in computing the partition function.

As the representation basis for this new Hilbert space, we take simultaneous eigenfunctions of the $z$-components of all the Pauli matrices. We designate these basis vectors with the symbol $\boldsymbol{n}_1$. The symbol $\boldsymbol{n}_1$, therefore, stands for a set of $2SN$ one bit variables where $N$ is the total number of lattice points in the original lattice. Then, the partition function of the original problem can be written as

$$Z = \sum_{\boldsymbol{n}_1} \left\langle \boldsymbol{n}_1 \left| \hat{P} e^{-\beta \mathcal{H}} \hat{P} \right| \boldsymbol{n}_1 \right\rangle. \tag{2.3}$$

Here, $\hat{P}$ is the projection operator to the subspace in which $\boldsymbol{S}_i^2 = S(S+1)$ for all $i$.

In order to evaluate each term in (2.3), we have to compute the matrix elements of the Boltzmann operator multiplied by the projection operator. This task is, however, practically impossible for large systems. Therefore, we usually use Suzuki-Trotter decomposition [7] to map the problem into a classical problem. Accordingly, the lattice we will work on is not the original lattice on which the quantum problem is defined. Instead, we will consider many layers, each of which is geometrically equivalent to the original lattice, and will take this set of layers as a hyper-lattice which has dimension one greater than the original dimension. In what follows we call the hyper-lattice simply the lattice. We specify a lattice point in this hyper-lattice by a set of two indices, e.g., $(k, i)$ where $i$ specifies the lattice point in the original lattice and $k$ specifies the layer to which the point belongs. Since $2SN$ one-bit variables are defined on each layer, the state of the system is described by $2SMN$ one-bit variables where $M$ is the number of layers. To make it easier to visualize the situation, we imagine that each of these one-bit variables is defined on a *vertex*. In other words, $2S$ vertices are associated with each lattice point. We specify a vertex by three indices, e.g., $(k, i, \mu)$. We write the whole set of variables as $\boldsymbol{n}$. The hyper-lattice can also be viewed as a collection of "shaded" plaquettes. Here a plaquette is a set of four lattice points $(k, i)$, $(k, j)$, $(k+1, i)$ and $(k+1, j)$ where $i$ and $j$ are nearest neighbors in the original lattice. A plaquette is also a set of $8S$ vertices. The use of the word "shaded" originated in the fact that plaquettes on which the four body interactions are defined are shaded or hatched in almost all the previous pictorial representations of the hyper lattice to distinguish them from other plaquettes each of which is merely a set of four nearest neighbor lattice points. In what follows, we call a "shaded" plaquette simply a plaquette. Now, our problem can be written as

$$Z = \sum_{\boldsymbol{n}} \prod_{p} \mathrm{sgn}(\boldsymbol{n}(p)) w(\boldsymbol{n}(p)) \tag{2.4}$$

where the product is taken over the set of plaquettes and $\boldsymbol{n}(p)$ is the subset of $\boldsymbol{n}$ whose elements are included in the plaquette $p$. Namely, for a plaquette $p = \{(k, i), (k, j), (k+1, i), (k+1, j)\}$,

$$\begin{aligned}\boldsymbol{n}(p) = \{&n_{(k,i,1)}, n_{(k,i,2)}, \cdots, n_{(k,i,2S)}, n_{(k,j,1)}, n_{(k,j,2)}, \cdots, n_{(k,j,2S)},\\&n_{(k+1,i,1)}, n_{(k+1,i,2)}, \cdots, n_{(k+1,i,2S)}, n_{(k+1,j,1)}, n_{(k+1,j,2)}, \cdots, n_{(k+1,j,2S)}\}.\end{aligned} \tag{2.5}$$

The weight $w(\boldsymbol{n}(p))$ and $\text{sgn}(\boldsymbol{n}(p))$ are the absolute value and the sign of the local Boltzmann weight. Here, the local Boltzmann weight of the classical problem is a matrix element of an operator $\hat{P} e^{\hat{\Lambda}} \hat{P}$ where $\hat{\Lambda}$ is defined by

$$\hat{\Lambda} \equiv \sum_{\mu,\nu} \hat{\Lambda}_{\mu,\nu}, \tag{2.6}$$

$$\hat{\Lambda}_{\mu,\nu} \equiv K_0 + K_x \sigma^x_{i,\mu} \sigma^x_{j,\nu} + K_y \sigma^y_{i,\mu} \sigma^y_{j,\nu} + K_z \sigma^z_{i,\mu} \sigma^z_{j,\nu}. \tag{2.7}$$

The constants $K_\alpha$ ($\alpha = x, y, z$) in general depend on the plaquette and are related to the coupling constants $J_\alpha$ in such a way that the sum of $K_\alpha$'s for all plaquettes with which both $i$ and $j$ are associated equals $\beta J_\alpha$ where $\beta$ is the inverse temperature.

In the paper I, we argued that once we obtain a set of coefficients $v(g) \geq 0$ that satisfy

$$w(\boldsymbol{n}(p)) = \sum_{g \in \Gamma} v(g) \Delta(\boldsymbol{n}(p), g), \tag{2.8}$$

we can obtain a cluster algorithm. The resulting algorithm is characterized by the probability for the graph assignment

$$p(g|\boldsymbol{n}(p)) = v(g) \Delta(\boldsymbol{n}(p))/w(\boldsymbol{n}(p)). \tag{2.9}$$

In (2.8), $\Gamma$ is a set of graphs which depends on the magnitude of spins and the anisotropy of the model.

The graph $g$ is defined on the plaquette $p$ and the function $\Delta(\boldsymbol{n}(p), g)$ takes only two values 0 and 1. A graph consists of edges and vertices where an edge is an object which connects two vertices. Each edge has a color, green or red. Like vertices, edges are introduced here just for making visualization and description of the algorithm easier. The function $\Delta(\boldsymbol{n}(p), g)$ is defined as

$$\Delta(\boldsymbol{n}(p), g) \equiv \begin{cases} 1 & \text{(If every green edge connects vertices with the same value,} \\ & \text{and every red edge connects vertices with different values.)} \\ 0 & \text{(Otherwise)} \end{cases} \tag{2.10}$$

Now, the problem is to find a proper set of graphs $\Gamma$ and coefficients $v(g)$ that satisfy (2.8). Once we obtain these, the actual simulation goes as follows: Starting from an arbitrary initial state $\boldsymbol{n}$, we first assign a graph $g$ to each plaquette $p$ with the probability (2.9). Because of the definition of $\Delta$, a green edge can be assigned only to a pair of parallel Pauli spins whereas a red one can be assigned only to a pair of anti-parallel Pauli spins. When we finish this graph assignment for every plaquette, we view the union of all these graphs as a single global graph. Then in the next step, i.e., the flipping process, we flip each cluster in the global graph with probability $1/2$. These two steps, graph assignment and cluster flipping, constitute one Monte Carlo step of the cluster algorithm.

In the next section, we will discuss how we obtain $\Gamma$ and the solution $v(g)$ of the equation (2.8).

### III. THE DECOMPOSITION OF THE BOLTZMANN FACTOR

As we discussed in I, the basis for a cluster Monte Carlo algorithm is the decomposition of the local Boltzmann weight $w(\boldsymbol{n}(p))$ into a sum of terms each of which corresponds to a graph, namely, the right hand side of (2.8). In terms of operators, (2.8) is equivalent to decomposing the operator $\hat{\rho} \equiv |\hat{P} e^{\hat{\Lambda}} \hat{P}| = \hat{P} e^{|\hat{\Lambda}|} \hat{P}$ into the following form

$$\hat{\rho} = \sum_{g \in \Gamma} v(g) \hat{\Delta}(g) \quad (v(g) \geq 0) \tag{3.1}$$

where $\hat{\Delta}(g)$ is an operator whose matrix elements are $\Delta(\boldsymbol{n}(p), g)$ and $|\hat{X}|$ stands for the operator whose matrix elements are the absolute values of those of $\hat{X}$.

It is useful to consider a set of operators that can be written in the form similar to (3.1). We first define $B$ as a set of operators which correspond to graphs, i.e., $B \equiv \{\hat{\Delta}(g)|g \in \Gamma\}$. Then, we define $O(B)$ as a set of operators which are linear combinations of elements of $B$ with non-negative coefficients. It is obvious that this set is closed also with respect to the multiplication by a non-negative real numbers and the addition of two elements. It is closed also with respect to the multiplication of two elements. (In other words, we have to choose the basis set $B$ so that the set $O(B)$ is closed.) We can view $O(B)$ as a subset of the finite dimensional linear space spanned by the basis set $B$. Therefore, we can represent every operator in $O(B)$ as a finite dimensional vector. At the same time, multiplying an operator $\hat{X} \in O(B)$ by another operator $\hat{Y} \in O(B)$ from left can be viewed as some linear operation specified by $\hat{Y}$ applied to an operator (i.e., a vector) $\hat{X}$. Therefore, we can represent every operator in $O(B)$ also as a finite dimensional matrix. With these definitions, (3.1) is written as

$$\hat{\rho} = \sum_{\hat{X} \in B} v(\hat{X}) \hat{X} \quad (v(\hat{X}) \geq 0), \tag{3.2}$$

which gives the vector representation $\boldsymbol{v}$, whose elements are $v(\hat{X})$, for the operator $\hat{\rho}$. Here, we used the same symbol $v$ as in (3.1) for the vector elements, since there is one-to-one correspondence between graphs in $\Gamma$ and the basis operators in $B$. Note that (2.8), (3.1) and (3.2) are equivalent to each other.

In order to obtain this vector representation for $\hat{\rho}$, we first compute the vector representation of $|\hat{\Lambda}_{\mu,\nu}|$

$$|\hat{\Lambda}_{\mu,\nu}| = \sum_{\hat{X} \in B} a_{\mu,\nu}(\hat{X}) \hat{X} \quad (a_{\mu,\nu}(\hat{X}) \geq 0). \tag{3.3}$$

Since the operator $\hat{\Lambda}_{\mu,\nu}$ influences only two Pauli spins, $\sigma_{i,\mu}^{\alpha}$ and $\sigma_{j,\nu}^{\alpha}$, the problem is essentially the same as the $S = 1/2$ problem as far as the solution of (3.3) is concerned. It is sufficient to consider graphs defined on only four vertices $(k, i, \mu)$, $(k, j, \nu)$, $(k+1, i, \mu)$ and $(k+1, j, \nu)$ instead of graphs defined on $8S$ vertices. In other words, the solution of (3.3) is given by a direct product of the solution of the $S = 1/2$ problem and the identity operator for the dimensions related to neither one of the Pauli spins $\sigma_{i,\mu}^{\alpha}$ and $\sigma_{j,\nu}^{\alpha}$. Once we obtain (3.3), we can easily get the expression

$$|\hat{\Lambda}| = \sum_{\mu,\nu} |\hat{\Lambda}_{\mu,\nu}| = \sum_{\hat{X} \in B} a(\hat{X}) \hat{X} \quad (a(\hat{X}) \geq 0), \tag{3.4}$$

where $a(\hat{X}) = \sum_{\mu,\nu} a_{\mu,\nu}(\hat{X})$. Then, we can calculate the vector representation of $(\hat{\Lambda})^n$ ($n = 2, 3, 4, \cdots$) by operating $n-1$ times the matrix representation of $\hat{\Lambda}$ to the vector $\boldsymbol{a}$.

In this way, we can calculate the vector representation of the operator $e^{|\hat{\Lambda}|}$ up to any finite order in $|\hat{\Lambda}|$ by Taylor expanding $e^{|\hat{\Lambda}|}$. Since the radius of convergence is infinite in this case, we should be able to obtain a good approximation by truncating the Taylor series at some order. In addition, $\hat{P}$ also belongs to $O(B)$ and it is easy to obtain its vector and matrix representations. In fact, the vector representation is given by a simple formula

$$\hat{P} = \sum_{g \in \Pi} \frac{1}{((2S)!)^2} \hat{\Delta}(g) \tag{3.5}$$

where $\Pi$ is the set of graphs that consist of green vertical edges and have no vertex shared by more than one edge. In other words, $\Pi$ is the set of graphs which correspond to permutations of vertices. Therefore, the entire process of computing the vector representation of $\hat{\rho}$ can be done at least numerically. This procedure will be explained again in the next section with a concrete example.

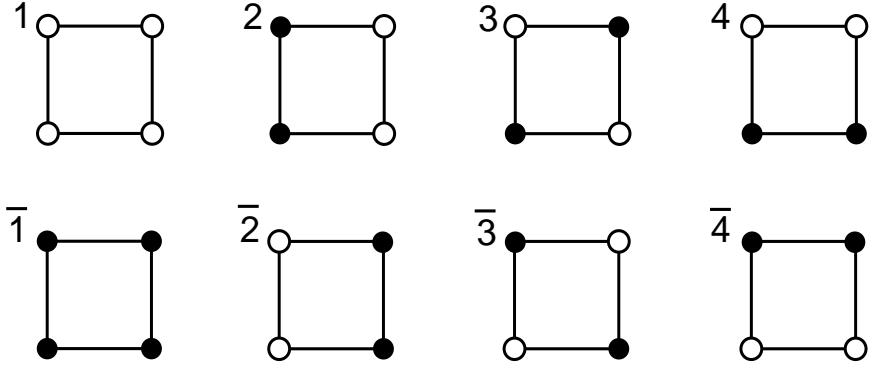

FIG. 1. Eight possible states of a plaquette for $S = 1/2$.

Now, our first problem is how to obtain the coefficient $a(g)$ in (3.3). As we discussed above, decomposition of $\hat{\Lambda}$ can be obtained through that of $\hat{\Lambda}_{\mu,\nu}$ which is essentially an $S = 1/2$ problem. Therefore, we consider an $S = 1/2$ problem with only two Pauli spins for which we do not need indices $\mu$ or $\nu$. Here we consider an operator given by

$$\hat{\Lambda} \equiv K_0 + K_x \sigma_i^x \sigma_j^x + K_y \sigma_i^y \sigma_j^y + K_z \sigma_i^z \sigma_j^z. \tag{3.6}$$

Since our Hilbert space in the present case is four dimensional, operators can naturally be expressed as $4 \times 4$ matrices. (Note, however, that this matrix representation is different from the one mentioned above.) Then, if we define

$$K_1 \equiv K_0 + K_z, \quad K_2 \equiv K_0 - K_z, \quad K_3 \equiv |K_x + K_y|, \quad \text{and} \quad K_4 \equiv |K_x - K_y|, \tag{3.7}$$

the matrix that represents $|\hat{\Lambda}|$ is

$$K \equiv \begin{pmatrix} K_1 & 0 & 0 & K_4 \\ 0 & K_2 & K_3 & 0 \\ 0 & K_3 & K_2 & 0 \\ K_4 & 0 & 0 & K_1 \end{pmatrix}. \tag{3.8}$$

We have assumed sufficiently large $K_0$ so that $K_1, K_2 \geq 0$. When the problem is mapped to a classical problem by the Suzuki-Trotter decomposition, each one of the matrix elements corresponds to a Boltzmann weight for a local state of a plaquette. From (3.8), it is obvious that only 8 among 16 states can have non-zero Boltzmann weights. These 8 states are shown in Fig. 1. If we number these states as shown in Fig. 1, the states $\sigma$ and $\bar{\sigma}$ correspond to the same matrix element $K_\sigma$ ($\sigma = 1, 2, 3, 4$). We should also notice that we can regard the matrix in (3.8) as the local weight for an 8-vertex model. Therefore, the graph decomposition presented below gives cluster algorithms for the 8-vertex model.

Now, we consider the local graph that we can use to decompose the Boltzmann operator. A local graph partially fixes relative orientations of spins. For example, two spins connected by a red bond point to opposite directions. After flipping clusters, wwo spins not connected by bonds can have either relative orientation, parallel or anti-parallel. In the case of the $XXZ$ model, because of particle number conservation, we could use only six types of local graphs, i.e., $G^{(\sigma\tau)}$ ($\sigma, \tau \neq 4$) in Fig. 2. This was because if we assigned a graph other than these four to a plaquette, the local configuration resulting from flipping a cluster is not guaranteed to satisfy particle number conservation. In the present case, instead of particle number conservation, the local configuration must satisfy a weaker condition, conservation of the parity of the particle number. It can be expressed as

$$m_{bl} + m_{br} \equiv m_{tl} + m_{tr} \mod 2. \tag{3.9}$$

Here, $m_{bl}$ is the sum $\sum_\mu m_{(k,i,\mu)}$ where $(k, i)$ is the bottom-left corner of the plaquette. Other integers $m_{br}, m_{tl}$ and $m_{tr}$ are defined in a similar fashion for the bottom-right, top-left and top-right corners, respectively. There are only ten local graphs that do not violate this parity conservation after flipping any set of clusters in the graph. These graphs are shown in Fig. 2. We denote these graphs as $G^{(\sigma\tau)}$ ($\sigma, \tau = 1, 2, 3, 4$) as indicated in Fig. 2. Namely, in this case,

$$\Gamma = \{G^{(\sigma\tau)} | \sigma, \tau = 1, 2, 3, 4; \sigma \leq \tau\} \quad \text{and} \quad B = \{\hat{\Delta}(G^{(\sigma\tau)}) | \sigma, \tau = 1, 2, 3, 4; \sigma \leq \tau\}. \tag{3.10}$$

In general, the graph $G^{(\sigma\tau)}$ can be assigned only to four states $\sigma, \tau, \bar{\sigma}$ and $\bar{\tau}$ (two states $\sigma$ and $\bar{\sigma}$ if $\sigma = \tau$). After flipping edges in the graph, the resulting state is one of these states. In other words,

$$\Delta(\xi, G^{(\sigma\tau)}) = \Delta(\bar{\xi}, G^{(\sigma\tau)}) = \begin{cases} 1 & (\xi = \sigma, \xi = \tau) \\ 0 & (\text{Otherwise}) \end{cases}, \tag{3.11}$$

or, in the matrix representation used for (3.8), $\hat{\Delta}(G^{(11)})$, $\hat{\Delta}(G^{(22)})$, $\hat{\Delta}(G^{(33)})$ and $\hat{\Delta}(G^{(44)})$ are represented by

$$\begin{pmatrix} 1 & 0 & 0 & 0 \\ 0 & 0 & 0 & 0 \\ 0 & 0 & 0 & 0 \\ 0 & 0 & 0 & 1 \end{pmatrix}, \begin{pmatrix} 0 & 0 & 0 & 0 \\ 0 & 1 & 0 & 0 \\ 0 & 0 & 1 & 0 \\ 0 & 0 & 0 & 0 \end{pmatrix}, \begin{pmatrix} 0 & 0 & 0 & 0 \\ 0 & 0 & 1 & 0 \\ 0 & 1 & 0 & 0 \\ 0 & 0 & 0 & 0 \end{pmatrix}, \text{ and } \begin{pmatrix} 0 & 0 & 0 & 1 \\ 0 & 0 & 0 & 0 \\ 0 & 0 & 0 & 0 \\ 1 & 0 & 0 & 0 \end{pmatrix}, \tag{3.12}$$

respectively. Note also that
$$\hat{\Delta}(G^{(\sigma\tau)}) = \hat{\Delta}(G^{(\sigma\sigma)}) + \hat{\Delta}(G^{(\tau\tau)}) \quad (\sigma \neq \tau). \tag{3.13}$$
Therefore, using (3.8), the equation (3.4) can be simplified as
$$K_\sigma = \sum_\tau a_{(\sigma\tau)} \quad (a_{(\sigma\tau)} \geq 0). \tag{3.14}$$
Here, $a_{(\sigma\tau)}$ is the abbreviation for $a(\hat{\Delta}(G^{(\sigma\tau)}))$. We have identified $(\sigma\tau)$ with $(\tau\sigma)$. By expressing $K_\sigma$ as a column vector $\boldsymbol{K}$ and $a_{(\sigma\tau)}$ as a matrix $A$, we can rewrite (3.14) as
$$\boldsymbol{K} = A\boldsymbol{1}, \tag{3.15}$$
where $\boldsymbol{1}$ is the four dimensional vector whose elements are all 1. Our problem is thus reduced to a problem of finding the $4 \times 4$ matrix $A$ satisfying (3.15) given a four dimensional vector $\boldsymbol{K}$ with the constraint that $a_{(\sigma\tau)} \geq 0$ and $a_{(\sigma\tau)} = a_{(\tau\sigma)}$.

In general, many solutions for this equation exist, although which solution gives the most effective algorithm has not been studied extensively. In I, we briefly described a procedure based on the maximum entropy method [8] for choosing a feasible solution when we have no reason for favoring one of them over the other. In what follows, we will see at least one meaningful solution exists for an arbitrary set of parameters, $K_\sigma$. We will also see that if and only if the largest among $K_\sigma$'s is not larger than the sum of all others, we can get a solution corresponding to a loop algorithm.

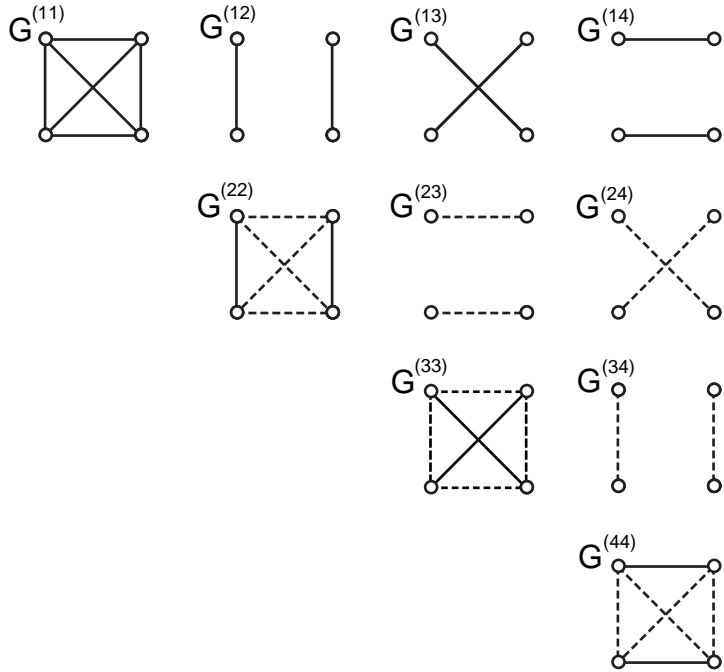

FIG. 2. Ten possible graphs of a plaquette for $S = 1/2$.

In what follows, we consider the case $K_z \geq 0$ and $K_x K_y \geq 0$. In this case, by taking sufficiently large $K_0$, we have the inequality $K_1 \geq K_2 \geq K_3 \geq K_4 \geq 0$ because of the definition (3.7) of $K_\sigma$. Once we get a solution in this case, we can get a solution in other cases as well simply by permuting indices. This is possible because (3.14) is invariant with respect to the index permutation. We should also note that the trivial solution

$$a_{(\sigma\sigma)} = K_\sigma \ (\sigma = 1,2,3,4), \quad \text{and} \quad a_{\sigma\tau} = 0 \quad (\text{if } \sigma \neq \tau) \tag{3.16}$$

does not yield any useful algorithm because in this case all vertices in the system are connected each other to form a single cluster. In such a case, only two states, the initial state and the reversed state, can be realized. It is also argued in I that in general we should minimize the possibility of two vertices being connected. Therefore, in general, we should avoid a solution in which $a_{(\sigma\sigma)}$'s are unnecessarily large because $G^{(\sigma\sigma)}$ connects all four vertices of the plaquette and 'locks' them into a single degree of freedom whereas $G^{(\sigma\tau)}$ connects them only pairwise and leaves two degrees of freedom. For example, if we have a solution in which $a_{(11)} > a_{(22)} > 0$, we can get a better solution by increasing $a_{(12)}$ by $a_{(22)}$ and decreasing $a_{(11)}$ and $a_{(22)}$ by the same amount. (As a result, $a_{(22)}$ becomes zero.) In other words, we can get a better solution by replacing $G^{(11)}$ and $G^{(22)}$ by $G^{(12)}$. From this example, it is clear that in the best solution, more than one $a_{(\sigma\sigma)}$ cannot be non-zero.

With these conditions, we now consider the following three cases: 1) $K_2 + K_3 + K_4 \leq K_1$, 2) $K_2 + K_3 - K_4 \leq K_1 \leq K_2 + K_3 + K_4$, and 3) $K_1 \leq K_2 + K_3 - K_4$. Equivalently, in terms of $K_x, K_y$ and $K_z$, 1) $|K_z|$ is the largest among $|K_x|, |K_y|$ and $|K_z|$, 2) $|K_z|$ is the second largest among three, and 3) $|K_z|$ is the smallest among three.

In the case 1), obviously we cannot have a solution in which $a_{(11)}$ is zero because

$$0 < K_1 - K_2 - K_3 - K_4 < a_{(11)} - \sum_{\sigma,\tau=2,3,4} a_{(\sigma\tau)} < a_{(11)}. \tag{3.17}$$

This means that a loop algorithm solution does not exist. Considering the remark in the last paragraph, we should seek a solution in which $a_{(22)} = a_{(33)} = a_{(44)} = 0$. Among such solutions, the one that minimizes $a_{(11)}$ is

$$a_{(11)} = K_1 - K_2 - K_3 - K_4, \tag{3.18}$$
$$a_{(1\sigma)} = K_\sigma \quad (\text{for } \sigma = 2,3,4), \tag{3.19}$$
$$a_{(\sigma\tau)} = 0 \quad (\text{for } \sigma \neq 1 \text{ and } \tau \neq 1). \tag{3.20}$$

In the case 2) and 3), as we will see below, solutions exist for which all $a_{(\sigma\sigma)}$'s are zero. This kind of solution corresponds to a loop algorithm in which the clusters formed do not have any branching, i.e., they are merely loops. In what follows, we will consider only such solutions for the motivation described above, although other solutions also exist. Given this condition, there are only 6 independent variables to be determined with the 4 independent equations (3.14). Therefore, the solution space is in general two-dimensional. To be more specific, in the matrix form, the general solution of (3.15) for the case 2) and 3) must have the following form:

$$A = A_0 + uU + vV \tag{3.21}$$

where $A_0$ is a special solution of (3.15), $u$ and $v$ are real numbers, and $U$ and $V$ are symmetric matrices which satisfy $U\mathbf{1} = V\mathbf{1} = \mathbf{0}$. To be specific,

$$U \equiv \begin{pmatrix} 0 & -1 & 0 & 1 \\ -1 & 0 & 1 & 0 \\ 0 & 1 & 0 & -1 \\ 1 & 0 & -1 & 0 \end{pmatrix} \quad \text{and} \quad V \equiv \begin{pmatrix} 0 & -1 & 1 & 0 \\ -1 & 0 & 0 & 1 \\ 1 & 0 & 0 & -1 \\ 0 & 1 & -1 & 0 \end{pmatrix}. \tag{3.22}$$

It is easy to see that

$$A_0 \equiv \frac{1}{2} \begin{pmatrix} 0, & 2K_2, & K_1 - K_2 + K_3 - K_4, & K_1 - K_2 - K_3 + K_4 \\ 2K_2, & 0, & 0, & 0 \\ K_1 - K_2 + K_3 - K_4, & 0, & 0, & -K_1 + K_2 + K_3 + K_4 \\ K_1 - K_2 - K_3 + K_4, & 0, & -K_1 + K_2 + K_3 + K_4, & 0 \end{pmatrix}, \tag{3.23}$$

satisfies (3.15). The condition $a_{(\sigma\sigma)} \geq 0$ imposes a restriction on the range of $(u, v)$. In the case 2), we have

$$0 \leq u, \quad 0 \leq v, \quad u + v \leq \frac{1}{2}(-K_1 + K_2 + K_3 + K_4). \tag{3.24}$$

In the case 3), if we define $A_1 \equiv A_0 + (1/2)(-K_1 + K_2 + K_3 + K_4)U$ and $u' \equiv u - (1/2)(-K_1 + K_2 + K_3 + K_4)$, the general solution can be written as

$$A = A_1 + u'U + vV. \tag{3.25}$$

The condition on $(u', v)$ is

$$0 \leq u', \quad 0 \leq v, \quad u' + v \leq K_4. \tag{3.26}$$

Thus, we have obtained the whole set of loop algorithm solutions of (3.15).

## IV. AN EXAMPLE — THE $XY$ MODEL WITH THE QUANTIZATION AXIS IN THE EASY PLANE

### A. The algorithm

In this section, we discuss how the decomposition of the operator $\hat{\Lambda}$ (3.4) presented in the last section is used for constructing a loop algorithm. As an example, we take the $XY$ model with the quantization axis lying in the easy plane. This is equivalent to choosing $J_x = J_z = J \geq 0$ and $J_y = 0$ while we use the conventional representation of Pauli matrices in which $z$-components are diagonal matrices. This representation is useful not only for giving an example for what we have discussed but also for some practical purposes. Namely, with this representation, we can easily calculate the correlations between spin-components in the easy-plane, i.e., $z$-components in the present case. Since the singularity due to the Kosterlitz-Thouless transition is manifested most strongly in such correlations, taking this representation is advantageous. On the other hand, we have to develop an algorithm different from the one for the 6-vertex model [5], because with this representation the system no longer maps to a 6-vertex model even in the $S = 1/2$ case.

Obviously, the $XY$ model is a marginal case that belongs to both case 1) and case 2) described in the last section. The solution is given by

$$
\begin{aligned}
a_{(12)} &= K_2 = K_0 - K_z = K_0 - K, \\
a_{(13)} &= K_3 = |K_x + K_y| = K, \\
a_{(14)} &= K_4 = |K_x - K_y| = K, \\
a_{(\sigma\tau)} &= 0 \quad (\text{if } (\sigma\tau) = (11), \text{ or, } \sigma \neq 1 \text{ and } \tau \neq 1).
\end{aligned}
\tag{4.1}
$$

Namely,

$$|\hat{\Lambda}| = (K_0 - K)\hat{\Delta}_2 + K\hat{\Delta}_3 + K\hat{\Delta}_4, \tag{4.2}$$

where $\hat{\Delta}_\sigma$ is an abbreviation for $\hat{\Delta}(G^{(1\sigma)})$. Note that the linear space spanned by $\hat{\Delta}_2$, $\hat{\Delta}_3$ and $\hat{\Delta}_4$ is closed with respect to the multiplication. Therefore, as the basis set $B$, we can take $\{\hat{\Delta}_2, \hat{\Delta}_3, \hat{\Delta}_4\}$. To be more specific,

$$
\begin{aligned}
\hat{\Delta}_2\hat{\Delta}_2 &= \hat{\Delta}_2, & \hat{\Delta}_2\hat{\Delta}_3 &= \hat{\Delta}_3, & \hat{\Delta}_2\hat{\Delta}_4 &= \hat{\Delta}_4, \\
\hat{\Delta}_3\hat{\Delta}_2 &= \hat{\Delta}_3, & \hat{\Delta}_3\hat{\Delta}_3 &= \hat{\Delta}_2, & \hat{\Delta}_3\hat{\Delta}_4 &= \hat{\Delta}_4, \\
\hat{\Delta}_4\hat{\Delta}_2 &= \hat{\Delta}_4, & \hat{\Delta}_4\hat{\Delta}_3 &= \hat{\Delta}_4, & \hat{\Delta}_4\hat{\Delta}_4 &= 2\hat{\Delta}_4.
\end{aligned}
\tag{4.3}
$$

Note that

$$\hat{\Delta}_\sigma \hat{\Delta}_\tau = \hat{\Delta}_\tau \hat{\Delta}_\sigma. \tag{4.4}$$

Using (4.3) and (4.4), we have

$$
\begin{aligned}
e^{\hat{\Lambda}} &= e^{(K_0-K)\hat{\Delta}_2} e^{K\hat{\Delta}_3} e^{K\hat{\Delta}_4} \\
&= [e^{(K_0-K)}\hat{\Delta}_2][\cosh K \hat{\Delta}_2 + \sinh K \hat{\Delta}_3][\Delta_2 + \frac{1}{2}(e^{2K} - 1)\hat{\Delta}_4] \\
&= e^{K_0}(e^{-K}\cosh K \hat{\Delta}_2 + e^{-K}\sinh K \hat{\Delta}_3 e^K \sinh K \hat{\Delta}_4).
\end{aligned}
\tag{4.5}
$$

This means that

$$v_{(12)} = e^{-K}\cosh K, \quad v_{(13)} = e^{-K}\sinh K, \quad v_{(14)} = e^K \sinh K \tag{4.6}$$

in the equation (2.8). (We omitted $e^{K_0}$ since it does not affect the resulting algorithm at all.) Therefore, the equation (2.9) leads to

$$p((\sigma\tau)|\tau) = p((\sigma\tau)|\bar{\tau}) = \frac{v_{(\sigma\tau)}}{\sum_{\sigma'} v_{(\sigma'\tau)}}, \tag{4.7}$$

where $p((\sigma\tau)|\xi)$ stands for $p(G^{(\sigma\tau)}|\xi)$. More explicitly,

$$
\begin{aligned}
p((12)|1) &= e^{-2K}, & p((13)|1) &= e^{-2K}\tanh K, & p((14)|1) &= \tanh K, \\
p((12)|2) &= 1, & p((13)|3) &= 1, & p((14)|4) &= 1,
\end{aligned}
\tag{4.8}
$$

$$p((\sigma\tau)|\xi) = 0 \quad (\text{if } \xi \neq \sigma \text{ and } \xi \neq \tau) \tag{4.9}$$

and

$$p((\sigma\tau)|\bar{\xi}) = p((\sigma\tau)|\xi). \tag{4.10}$$

For the systems with spin $S$ larger than $1/2$, we can follow essentially the same line to obtain the labeling probability. In other words, first we find a set of basis operators $B$ that spans a linear space closed with respect to the multiplication, then calculate the vector representation of $e^{|\hat{A}|}$ in terms of these basis operators. However, we cannot in general simplify this calculation as we did in the case of $S = 1/2$, because commutativity (4.4) does not hold for the basis operators in the case of $S > 1/2$. We can still at least numerically calculate the expansion of $e^{|\hat{A}|}$ as discussed in I by Taylor series expanding $e^{|\hat{A}|}$ in terms of $|\hat{A}|$. Since the radius of convergence circle of this Taylor expansion is infinity, we should be able to get a good approximation if we truncate the series at the finite but sufficiently high order.

For example, in the case where $S = 1$, $J_x = J_z = J \geq 0$ and $J_y = 0$, similar to $S = 1/2$ case, only graphs we have to take into account are those which connect vertices pair-wise. Therefore, there are $8!/(2^4 4!) = 105$ distinct graphs to consider. In other words, 105 operators that corresponds to these graphs constitute the basis set $B$. The product of arbitrary two basis operators is another basis operator except for the numerical factor $2^m$ due to inner closed loops as discussed in I. Namely, for arbitrary two elements $\hat{X}$ and $\hat{Y}$ of $B$, another element $\hat{Z}(\hat{X}, \hat{Y})$ of $B$ and an integer $m(\hat{X}, \hat{Y})$ exist that satisfy

$$\hat{X}\hat{Y} = 2^{m(\hat{X},\hat{Y})}\hat{Z}(\hat{X}, \hat{Y}). \tag{4.11}$$

It is tedious but straight-forward to calculate $m(\hat{X}, \hat{Y})$ and $\hat{Z}(\hat{X}, \hat{Y})$ for all possible pairs of $\hat{X}$ and $\hat{Y}$ and prepare a table similar to (4.3). Once we have this table, we can calculate the product of two arbitrary operators in the linear space $O(B)$, easily. To be specific, for two operators in $O(B)$, $\hat{S} \equiv \sum_{\hat{X} \in B} s(\hat{X})\hat{X}$ and $\hat{T} \equiv \sum_{\hat{X} \in B} t(\hat{X})\hat{X}$, we have

$$\hat{S}\hat{T} = \sum_{\hat{X},\hat{Y}} s(\hat{X})t(\hat{Y})\hat{X}\hat{Y} = \sum_{\hat{X},\hat{Y}} s(\hat{X})t(\hat{Y}) 2^{m(\hat{X},\hat{Y})} \hat{Z}(\hat{X}, \hat{Y}) = \sum_{\hat{X}} u(\hat{X})\hat{X} \tag{4.12}$$

where

$$u(\hat{W}) \equiv \sum_{\substack{\hat{X},\hat{Y} \\ \hat{W} = \hat{Z}(\hat{X},\hat{Y})}} s(\hat{X})t(\hat{Y}) 2^{m(\hat{X},\hat{Y})}. \tag{4.13}$$

In other words, multiplying $\hat{T}$ by $\hat{S}$ from left is equivalent to multiplying a vector defined by

$$(\boldsymbol{T})_{\hat{X}} \equiv t(\hat{X}) \tag{4.14}$$

by a matrix

$$(S)_{\hat{X},\hat{Y}} \equiv \sum_{\substack{\hat{W} \\ \hat{X} = \hat{Z}(\hat{W},\hat{Y})}} s(\hat{W}) 2^{m(\hat{W},\hat{Y})} \tag{4.15}$$

from left. Therefore, since $\hat{P}$ and $|\hat{\Lambda}|$ is an element of $O(B)$, calculating the vector representation of $\hat{P}|\hat{\Lambda}|^n\hat{P}$ in terms of the basis operators is computationally straight-forward. Hence, $\hat{P}e^{|\hat{\Lambda}|}\hat{P}$ can be calculated at least numerically. In fact, the calculation described here can be simplified significantly if we take symmetry into account.

There is yet another way of calculating $\hat{P}e^{|\hat{\Lambda}|}\hat{P}$. Namely, explicitely solving the linear algebraic equation (2.8) with respect to $v(g)$. It is obvious the solution obtained by the former method is unique and satisfies (2.8), although the solution of (2.8) is not necessarily unique.

### B. Comparison of the conventional algorithm and the loop algorithm

We applied the loop algorithm described in the last subsection to the $XY$-model on a square lattice with the periodic boundary condition. At the same time, we applied the conventional algorithm to the same system to compare the efficiency of the two algorithms. Ding and Makivić [9] reported that this system undergoes a Kosterlitz-Thouless type phase transition at $T \sim 0.35$. Therefore, we expect critical-slowing down for the conventional algorithm. We also expect another slowing-down as the imaginary time spacing becomes smaller with a fixed temperature as it happened [4] in the one-dimensional $S = 1$ antiferromagnetic Heisenberg model.

The details of the conventional algorithm used in this paper is presented in Appendix. The conventional algorithm is the same as the one used in [10] except that we included "diagonal" flips to make the simulation ergodic. (As we will see in the Appendix, the algorithm in [10] is not completely ergodic although the effect of this non-ergodicity may not be significant.) We remark here that one usually needs to be concerned about the ergodicity in the conventional algorithm and that the actual computer programs for the conventional algorithm tend to be complicated because to ensure ergodicity one has to incorporate several different updates in a non-unified fashion. If we have four different kinds of flips, we usually write four different subroutines. On the other hand, the cluster algorithm is less likely non-ergodic because in most cases the cluster algorithm includes wider class of updates than the conventional algorithm. As discussed in I, we can even prove ergodicity for the cluster algorithm in some cases. In addition, in a cluster algorithm, all kinds of updates can be realized in a unified fashion in actual computer programs.

We calculated the integrated auto-correlation time defined [11] by

$$\tau_X(b) \equiv \frac{bv_X(b)}{2v_X(1)} \qquad (4.16)$$

where $v_X(b)$ is the variance of the distribution of the bin averages with the bin length of $b$. This quantity should be equal to the integrated auto-correlation time defined by

$$\tau_X^{(int)} \equiv \sum_{t=0}^{\infty} \langle X(\tau)X(0) \rangle_{\text{MC}} / \langle X(0)X(0) \rangle_{\text{MC}} \qquad (4.17)$$

in the limit of $b \to \infty$. Here, $X(t)$ is an arbitrary physical quantity measured at the $t$-th Monte Carlo step. As a function of increasing $b$, $\tau_X(b)$ is generally a non-decreasing function. Upto a certain point, say, $b \sim \tau_X^{(cor)}$, $\tau_X(b)$ increases and after this point, roughly speaking, it takes the constant value $\tau_X^{(int)}$. In a typical simulation that we have done, $\tau_X^{(int)} \ll \tau_X^{(cor)}$. We regard $\tau_X^{(cor)}$ as the number of Monte Carlo steps needed for decorrelating two measurements of $X$ completely. In order to obtain an estimate of $\tau_X(b)$ with a small statistical error, we had to perform a simulation much longer than $b$, in general. Therefore, in case the total number of Monte Carlo steps $T$ is larger but not much larger than $\tau_X^{(cor)}$, it is difficult to judge if the function $\tau_X(b)$ has reached the plateau because the plateau is blurred by large statistical errors. Hence, we often cannot estimate $\tau_X^{(int)}$ precisely. In such cases, we took $\tau_X(b)$ at the largest bin length $b$ where a statistically precise estimate is still possible and regarded it as a lower bound of $\tau_X^{(int)} \equiv \tau_X(\infty)$. We calculated $\tau_X(b)$ with magnetization ($M \equiv \sum_{\boldsymbol{R}} S_z(\boldsymbol{R})$), susceptibility ($M^2/N$), and nearest-neighbor spin-spin correlations ($\sum_{\boldsymbol{R}} S_\alpha(\boldsymbol{R}) S_\alpha(\boldsymbol{R}+\boldsymbol{\delta})/N$ ($\alpha = x, y, z$)).

As for the conventional algorithm, we found that the autocorrelation time of the nearest-neighbor spin-spin correlations for the in-plane spin components (i.e., $x$ and $z$ components) is equal to or larger than that for the susceptibility. On the other hand, the auto-correlation time for the $y$ components is smaller than that for the susceptibility in most cases we studied. In Table I, we show the autocorrelation times of the uniform magnetization and the magnetic susceptibility for the conventional algorithm.

| $\Delta\tau$ | $\beta$ | Magnetization | | | Susceptibility | | |
|---|---|---|---|---|---|---|---|
| | | $L=4$ | $L=8$ | $L=16$ | $L=4$ | $L=8$ | $L=16$ |
| 1.000 | 1 | 0.534(11) | 0.499(07) | 0.511(05) | 0.550(10) | 0.504(06) | 0.498(06) |
| 1.000 | 2 | 10.38(60) | 27.8(17) | 49.7(LB) | 5.67(51) | 9.60(68) | 20.3(LB) |
| 1.000 | 4 | 32.7(26) | 139.6(54) | 496(LB) | 16.8(11) | 51.3(24) | 242(LB) |
| 0.500 | 1 | 0.768(23) | 0.700(31) | | 0.741(32) | 0.647(37) | |
| 0.500 | 2 | 10.6(15) | 24.8(14) | | 5.68(59) | 12.1(11) | |
| 0.500 | 4 | 51.1(LB) | | | 26.6(16) | | |
| 0.250 | 1 | 0.745(30) | 0.678(20) | | 0.757(09) | 0.633(24) | |
| 0.250 | 2 | 13.26(36) | 28.70(86) | | 10.89(22) | 35.8(33) | |
| 0.250 | 4 | 115.4(52) | | | 53.2(24) | | |
| 0.125 | 1 | 0.727(08) | 0.725(25) | | 1.079(62) | 0.662(10) | |
| 0.125 | 2 | 14.93(67) | 30.7(23) | | 30.6(11) | 115(LB) | |
| 0.125 | 4 | 231(LB) | | | 144.0(84) | | |

TABLE I. The integrated autocorrelation times of the conventional algorithm for the uniform magnetization and the susceptibility. The system size is $L \times L$. The figures in the parentheses are statistical errors (one standard deviation). "LB" indicates that the value shown is the lower bound.

As we can clearly see, the correlation times for both the quantities show the slowing down as the temperature becomes low. It is also clear that for lower temperatures ($\beta = 2, 4$), the autocorrelation time grows fast as the system becomes larger. We consider this growth a finite size effect for $\beta = 2$ since this temperature is higher than the critical temperature $\beta_{\text{KT}} \sim 2.9$. Since $\beta = 4$ is below the critical temperature, we would observe the growth of the correlation time for larger systems. The slowing-down due to the increment of the Trotter number is also observed. But, it is significant only in the case of $\beta = 2$ and 4. On the other hand, we observed no clear evidence for any slowing-down in the case of the loop algorithm. The autocorrelation times for the magnetization are 0.5, as they should be, with statistical errors of a few percents. The autocorrelation times for the susceptibility are also almost constant and are around 1.0 or less (See Table II).

We should stress that the cluster algorithm has another significant advantage besides the reduction of the autocorrelation times, namely, the improved estimators [1,13]. In this paper, we calculated susceptibility by using an improved estimator. It is well-known that for the Ising model the improved estimator for the magntic susceptibility is simply the average cluster size. In the present case, too, the improved estimator of the magnetic susceptibility for $z$-components is proportional to the average cluster size. To be more specific,

$$\left\langle M_z^2 \right\rangle = \frac{1}{M^2} \left\langle \sum_c V_c^2 \right\rangle. \tag{4.18}$$

| $\Delta\tau$ | $\beta$ | $L = 4$ | $L = 8$ | $L = 16$ |
|---|---|---|---|---|
| 1.000 | 1 | 0.883(29) | 0.661(23) | 0.529(12) |
| 1.000 | 2 | 1.004(37) | 1.093(44) | 1.251(84) |
| 1.000 | 4 | 0.919(29) | 0.952(64) | 0.978(30) |
| 0.500 | 1 | 0.901(40) | 0.589(23) | |
| 0.500 | 2 | 1.033(37) | 1.077(61) | |
| 0.500 | 4 | 0.920(30) | | |
| 0.250 | 1 | 0.772(25) | 0.603(15) | |
| 0.250 | 2 | 0.973(40) | 1.085(40) | |
| 0.250 | 4 | 0.986(32) | | |
| 0.125 | 1 | 0.800(17) | 0.629(22) | |
| 0.125 | 2 | 0.991(29) | 1.138(50) | |
| 0.125 | 4 | 0.968(26) | | |

TABLE II. The integrated autocorrelation times of the loop algorithm for the susceptibility.

| $\beta$ | $L$ | Algorithm | $N_{MCS}$ | $N_{int}$ | $\langle M_z^2 \rangle /N$ | Error |
|---|---|---|---|---|---|---|
| 1 | 4 | C | 8,192 | 1 | 3.750 | 0.011 |
| 1 | 4 | L | 8,192 | 1 | 3.762 | 0.018 |
| 1 | 4 | I | 8,192 | 1 | 3.781 | 0.008 |
| 1 | 8 | C | 8,192 | 1 | 4.207 | 0.021 |
| 1 | 8 | L | 8,192 | 1 | 4.196 | 0.016 |
| 1 | 8 | I | 8,192 | 1 | 4.193 | 0.007 |
| 1 | 16 | C | 8,192 | 1 | 4.214 | 0.019 |
| 1 | 16 | L | 8,192 | 1 | 4.205 | 0.018 |
| 1 | 16 | I | 8,192 | 1 | 4.202 | 0.004 |
| 2 | 4 | C | 32,768 | 4 | 7.720 | 0.030 |
| 2 | 4 | L | 8,192 | 1 | 7.726 | 0.024 |
| 2 | 4 | I | 8,192 | 1 | 7.712 | 0.012 |
| 2 | 8 | C | 32,768 | 2 | 22.779 | 0.158 |
| 2 | 8 | L | 8,192 | 1 | 22.621 | 0.086 |
| 2 | 8 | I | 8,192 | 1 | 22.630 | 0.055 |
| 2 | 16 | C | 32,768 | 4 | 57.588 | 0.473 |
| 2 | 16 | L | 8,192 | 1 | 57.135 | 0.265 |
| 2 | 16 | I | 8,192 | 1 | 57.075 | 0.153 |
| 4 | 4 | C | 32,768 | 2 | 7.931 | 0.059 |
| 4 | 4 | L | 8,192 | 1 | 7.927 | 0.027 |
| 4 | 4 | I | 8,192 | 1 | 7.927 | 0.014 |
| 4 | 8 | C | 262,144 | 16 | 28.534 | 0.148 |
| 4 | 8 | L | 8,192 | 1 | 28.866 | 0.091 |
| 4 | 8 | I | 8,192 | 1 | 28.817 | 0.050 |
| 4 | 16 | C | 262,144 | 32 | 103.170 | 1.404 |
| 4 | 16 | L | 8,192 | 1 | 103.161 | 0.362 |
| 4 | 16 | I | 8,192 | 1 | 103.334 | 0.225 |

TABLE III. The estimated values for the susceptibility and the statistical error in the case of $\Delta\tau = 1$. In each entry, the top, middle and bottom figures are the estimates by the conventional algorithm (C), the loop algorithm (L), and the loop algorithm with the improved estimator (I), respectively. Each simulation consists of 10 sets where one set consists of $N_{MCS}$ Monte Carlo steps for measurements with a sufficiently large number of additional Monte Carlo steps for equillibration. A measurement of susceptibility is done every $N_{int}$ Monte Carlo steps. The last column is the estimate of statistical error in one standard-deviation.

In Table III, we listed the three sets of estimates for the susceptibility, i.e., the ones obtained with the conventional algorithm, the ones with the loop algorithm, and the ones with the loop algorithm and the improved estimator. For smaller values of $\Delta\tau$, the magnitude of the statistical errors for the conventional algorithm relative to that for the loop algorithm without the improved estimator tends to be larger. We can see that the difference between the statistical errors for the conventional algorithm and those for the loop algorithm without the improved estimator is consistent with the estimated integrated autocorrelation times shown in Table I and II. We also see that the improved estimator further reduces the statistical error considerably. Since using the improved estimator is equivalent to averaging over $2^{N_c}$ different configurations where $N_c$ is the number of clusters, its advantage is more significant when $N_c$ is larger, i.e., at higher tempereratures. On the other hand, the reduction of the auto-correlation time by using the loop algorithm is less significant at higher temperatures. Therefore, the two advantages of the loop algorithm, i.e., the reduction in the correlation times and the reduction in the variance of the thermodynamic distribution by the improved estimators, are complementary to each other.

The difference in the overall efficiency is striking. For example, in the case of $\Delta\tau = 1.0$, $\beta = 4$ and $L = 16$, the simulation by the conventional algorithm is 32 times longer than the loop algorithm simulation. However, the conventional algorithm yields a statistical error 6 times larger than that of the loop algorithm simulation with the improved estimator. Since the statistical error is proportional to the reciprocal of the square-root of the total number of the Monte Carlo steps, this difference in the efficiency is roughly equivalent to a factor $32 \times 6^2/r \sim 10^3/r$ in terms of the computational time where $r$ is the computaional time per one Monte Carlo step of the loop algorithm devided by that of the conventional algorithm. The factor $r$ strongly depends on the detail of the software and the hardware. In our particular case, $2 < r < 4$. Therefore, even for the small systems studied here, the difference is more than two orders of magnitude in the real CPU time.

## V. CONCLUSIONS

In this paper, we presented how to construct a cluster algorithm for a model described by the $XYZ$ Hamiltonian. The algorithm for the special case of $S = 1/2$ can be viewed as a cluster algorithm for the 8-vertex model as well as the algorithm for the quantum spin systems. The efficiency of the new algorithm is examined for the $S = 1/2$ $XY$ model on a square lattice. It is found that the integrated autocorrelation times of the new algorithm for various physical quantities do not show any significant slowing-down whereas the conventional algorithm suffers from slowing-down due to the low temperature and the small imaginary time spacing. Even for the small system sizes ($L \leq 16$) studied in this paper, the difference between the autocorrelation times for the two algorithms can be three orders of magnitude, and this difference is very likely much larger for larger systems. In addition, we observed that we can further reduce the statistical error by measuring quantities through improved estimators.

**Acknowledgment**

The auther would like to thank J. E. Gubernatis for useful comments and critical reading of the manuscript.

## APPENDIX: THE CONVENTIONAL ALGORITHM

The conventional algorithm used in this paper is the same as the one in [10] except that we added so-called "diagonal" loop flips. The algorithm in [10] consists of three types of updates; 1) local "space" flips, 2) local "time" flips and 3) global flips in the time direction. These flips, except for the diagonal flips, are described in [12]. It seems that in the simulation described in [10], the second global flip in [12], i.e., global flips in the space directions were not performed. In the case of the Heisenberg model, these global spatial flips are needed to make the algorithm ergodic. In other words, the other three types of updates do not change the total winding number of the world lines. Therefore, the simulation without the spatial global flips is not ergodic. In the case of the $XYZ$ model, the local conservation rule of the magnetization does not hold. It means that the worldlines are no longer well-defined. Therefore, it is not straight-forward to see if there are conserved quantities. In fact, however, conserved quantities exist if we do not include the spatial global flips or something equivalent to it. To see this, let us cut the system by a plane parallel to the $y$ and $t$ axes and consider the cross-section. Then, we take a half of the lattice points on this cross-section whose time coordinates $k$ are odd (see Fig. 3).

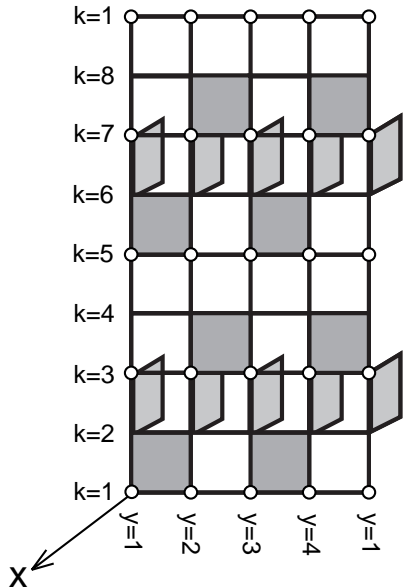

FIG. 3. A cross-section by a plane parallel to the $yz$-plane. The set $H$ is defined as the set of lattice points marked by open circles in the figure.

We refer to the set of these lattice points as $H$. Now, we define $P_x$ as the parity of the total number of lattice points in $H$ on which the spins are 'up' (i.e., $n_{(k,i)} = 1$). The $P_x$ does not depend on the location of the plane to cut the system. We claim that $P_x$ is not changed by local spatial flips, local temporal flips, or global temporal flips. The reason is simply that the intersection of $H$ and the set of lattice points that are affected by one of those flips always consists of an even number of lattice points. Since we can define $P_y$ in a similar fashion, our claim implies that there are at least two conserved numbers for the whole system in 2+1 dimensional systems.

Of course, the above definition of $P_x$ and $P_y$ is valid also for $XXZ$ models for which the local conservation rule for the magnetization applies. It is easy to see that $P_\alpha$ ($\alpha = x, y$) equals the parity of the total winding number in the $\alpha$-direction in the case of the $XXZ$ model. The above argument can be easily generalized to other dimensions and different Suzuki-Trotter decompositions by changing the definition of $H$ appropriately.

In this paper, we used global diagonal flips instead of the global spatial flips to make the algorithm ergodic. To be specific, a diagonal flip in the $x$-direction is a flip of a loop which is constructed by the following rules. 1) Take a lattice point $(x, y, t)$ where $t = 1, 2, \cdots, M$ specifies the imaginary time coordinate. 2) If $(x, y, t)$, $(x+1, y, t)$, $(x+1, y, t+1)$ and $(x, y, t+1)$ are four corners of a "shaded" plaquette, take $(x+1, y, t+1)$ as the next point. Otherwise, take $(x, y, t+1)$ instead. 3) Repeat 2) until the current $x$-coordinate coincides with the $x$-coordinate of the starting point. 4) Take $(x, y, t+1)$ as the next point. 5) Repeat 4) until the current point coincides with the starting point. In a similar fashion, we can define diagonal flips in the $y$-direction. In the actual simulations, for every Monte Carlo step, we included one set of diagonal flips in both $x$ and $y$ directions. Here, 'one set' of diagonal flips in the $x$-direction means the updates of all diagonal loops in the $x$-direction whose starting points are given by $(0, y, t)$ ($y = 1, 2, \cdots, L; t = 4, 8, 12, \cdots, M$). (Note that in the present case, the hyper-lattice is periodic with the period of 4 in the imaginary time direction.)

## REFERENCES

[1] P. W. Kasteleyn and F. M. Fortuin, J. Phys. Soc. Jpn. **26** Suppl., 11 (1969); C. M. Fortuin and P. W. Kasteleyn, Physica **57**, 536 (1972).
[2] R. H. Swendsen and J.-S. Wang, Phys. Rev. Lett. **58**, 86 (1987).
[3] N. Kawashima and J. E. Gubernatis, J. Stat. Phys., to appear.
[4] N. Kawashima and J. E. Gubernatis, Phys. Rev. Lett. **73**, 1295 (1994).
[5] H. G. Evertz, M. Marcu, and G. Lana, Phys. Rev. Lett. **70**, 875 (1993); H. G. Evertz and M. Marcu, in *Quantum Monte Carlo Method in Condensed Matter Physics*, edited by M.Suzuki, (World Scientific, Singapore, 1992), p.65.
[6] N. Kawashima, J. E. Gubernatis, and H. G. Evertz, Phys. Rev. B. **50** 136 (1994).
[7] M. Suzuki, Prog. Theor. Phys. **56**, 1454 (1976).
[8] M. Jarrell and J. E. Gubernatis, in preparation.
[9] H.-Q. Ding and M. S. Makivić, Phys. Rev. B **42**, 6827 (1990).
[10] H.-Q. Ding, Phys. Rev. B **45**, 230 (1992).
[11] M. P. Allen and D. J. Tildesley, *Computer Simulations of Liquids* (Oxford University Press, Oxford, 1987), Chap. 6.

[12] M. S. Makivić and H.-Q. Ding, Phys. Rev. B **43**, 3562 (1991).
[13] U. Wolff, Phys. Rev. Lett. **60**, 1461 (1988); U. Wolff, Nucl. Phys. **300[FS22]**, 501 (1988).